\documentclass[twocolumn,superscriptaddress,amsfont,amssymb,amsmath, showpacs,balancelastpage, nofootinbib]{revtex4-1}
\usepackage{graphicx,longtable,mathrsfs,color,array}
\usepackage[hidelinks]{hyperref}
\usepackage[usenames,dvipsnames]{xcolor}
\usepackage{amssymb,amsmath,mathtools,mathrsfs}
\usepackage{epsfig,subfigure,placeins,float}
\usepackage{booktabs,longtable,ctable,multirow}
\usepackage{exscale,relsize}
\usepackage[normalem]{ulem}
\usepackage{enumerate}
\usepackage{times, mathptmx}
\usepackage{colortbl} 
\usepackage[dvipsnames]{xcolor} 
\usepackage{pifont} 
\usepackage{amsmath} 

\allowdisplaybreaks

\begin{document}
 
\title{{Perturbations of a rotating black hole in DHOST theories}}
\author{Christos Charmousis}
\affiliation{Laboratoire de Physique Th\'eorique, CNRS, Univ.\ Paris-Sud, 
Universit\'e Paris-Saclay, 91405 Orsay, France}
\author{Marco Crisostomi}
\affiliation{Laboratoire de Physique Th\'eorique, CNRS, Univ.\ Paris-Sud, 
Universit\'e Paris-Saclay, 91405 Orsay, France}
\affiliation{Institut de physique th\'eorique, Univ.\ Paris Saclay, CEA, 
CNRS, 91191 Gif-sur-Yvette, France}
\affiliation{AIM, CEA, CNRS, Univ.\ Paris-Saclay, Univ.\ Paris Diderot,
Sorbonne Paris Cit\'e, F-91191 Gif-sur-Yvette, France}
\author{David Langlois}
\affiliation{AstroParticule et Cosmologie, UMR CNRS 7164, Universit\'e Paris-Diderot,
10 rue Alice Domon et L\'eonie Duquet, 75205 Paris, France}
\author{Karim Noui}
\affiliation{Institut Denis Poisson, UMR CNRS 7013, Universit\'e de Tours, Universit\'e d'Orl\'eans,
Parc de Grandmont, 37200 Tours, France}
\affiliation{AstroParticule et Cosmologie, UMR CNRS 7164, Universit\'e Paris-Diderot,
10 rue Alice Domon et L\'eonie Duquet, 75205 Paris, France}

 \date{\today}

\begin{abstract}
We study linear perturbations of a rotating black hole solution that has been recently discovered in degenerate higher-order scalar-tensor (DHOST) theories. We find a parametrization which permits the explicit resolution of the scalar perturbation while the  tensor perturbation is obtained as a Teukolsky equation supplemented by an effective source term.
The effective source term is related to the black hole hair and can be computed exactly for any value of the black hole spin.  
We discuss how the perturbations of the geometry and thus the emitted gravitational waves could be modified in comparison with general relativity.\

\end{abstract}

\maketitle

\section{Introduction}
 
The ability to detect gravitational waves (GW) has opened an extraordinary new window for cosmology and astrophysics.
It also offers the possibility of testing General Relativity (GR) directly and in the presence of very strong gravitational fields. 
From the latter perspective, 
the ringdown phase of a black hole merger is particularly interesting. Indeed, with the next generation of GW interferometers (such as the Einstein Telescope and LISA) it will be possible to measure multiple quasi-normal modes (QNM) emitted by the newly formed black hole and perform what is called {\it black hole spectroscopy} \cite{Berti:2016lat}.
This will permit to investigate potential deviations from the well-known QNM in GR \cite{Teukolsky:1973ha,Press:1973zz,Chandrasekhar:1985kt,Leaver:1985ax} (see \cite{Nollert:1999ji,Kokkotas:1999bd,Berti:2009kk} for reviews) and will provide a crucial test for gravity.

To fully exploit this new possibility, it is useful to explore what type of modifications could be expected from black holes in various models of modified gravity. Several no-hair theorems \cite{Bekenstein:1995un, Hui:2012qt} state that, under certain hypotheses, the only modification that can be obtained is at the level of perturbations \cite{Barausse:2008xv}, 
while the black hole solution is indistinguishable from GR. However, if some of the hypotheses are  relaxed,  differences can enter already at the background level: the black hole geometry can differ from Kerr \cite{Herdeiro:2014goa}, or a Kerr metric can be dressed with some non-trivial field  -- or ``hair", in which case we are in presence of the so called ``stealth'' solutions.

Several works have analysed perturbations and QNM of spherically symmetric black holes (with and without hair) 
in alternative theories of gravity, e.g. \cite{Kobayashi:2012kh,Kobayashi:2014wsa,Tattersall:2017erk,Tattersall:2018nve,Franciolini:2018uyq, Takahashi:2019oxz}. 
However, only few articles~\footnote{See \cite{Pani:2013ija} for Einstein-Maxwell theory, \cite{Suvorov:2019qow} for recent results in $f(R)$ theory and \cite{Tattersall:2018axd, Tattersall:2019pvx} for slowly rotating black holes without hair in Horndeski theories} deal with rotating black holes, 
one of the main reasons being the mere paucity of non-trivial rotating solutions in theories of modified gravity.

In the present work we consider a stealth rotating black hole solution obtained very recently \cite{Charmousis:2019vnf} in the context of Degenerate Higher Order Scalar Tensor (DHOST) theories \cite{Langlois:2015cwa,Crisostomi:2016czh, BenAchour:2016fzp}, which represent the most general class of covariant scalar-tensor theories with a single scalar degree of freedom (see \cite{Langlois:2018dxi} for a review).
This black hole solution is characterised by an exact Kerr (or, more generally, Kerr-de Sitter) geometry and a non-trivial scalar field which can be identified with families of geodesics of the spacetime itself~\footnote{The spherically symmetric versions of  \cite{Charmousis:2019vnf} were initially found for Horndeski theory in \cite{Babichev:2013cya}, whilst their disformed DHOST version and stability was investigated in \cite{Babichev:2017lmw} and \cite{Babichev:2018uiw}. A recent comprehensive review of hairy black hole solutions in scalar tensor gravity can be found in \cite{Lehebel:2018zga}.}.
For simplicity, in this paper we restrict ourselves to the solution of \cite{Charmousis:2019vnf} with no effective cosmological constant.

In the present work, we study the linear perturbations about this hairy black hole solution and obtain the tensor and scalar perturbation equation in a relatively simple form, so that one can
disentangle the scalar and  tensor perturbations. Remarkably, the tensor equation differs from the Teukolsky  equation  in GR only by the presence of an effective source term that depends on the ``hairy'' scalar perturbation, which can be explicitly solved.

\section{Perturbations of Kerr black holes in GR}

Before computing the relevant equations in our theory, it is useful to recall the main steps for the computation of gravitational perturbations of a Kerr black hole in general relativity. 
This problem is far more involved than the case of a  spherically symmetric black hole, due to the complexity of the geometry.
In Boyer-Lindquist coordinates, the Kerr metric reads 
\begin{eqnarray}
ds^2&= &-\frac{\Delta}{\rho^2} (dt - a \sin^2\theta \, d\varphi)^2 + \rho^2\left( \frac{dr^2}{\Delta} + d\theta^2\right) +\nonumber \\
&& +\frac{\sin^2\theta}{\rho^2} \left( a \, dt - (r^2+a^2) \, d\varphi\right)^2 \, ,
\label{Kerr}
\end{eqnarray}
where
\begin{eqnarray}
\Delta \equiv r^2+a^2-2Mr , \quad \rho^2 \equiv r^2 + a^2 \cos^2\theta \,.
\end{eqnarray}
$M$ is the mass and $a$ the angular momentum parameter ($a\leq M$) of the black hole.
Kerr's solution has an inner ($r=r_I$) and an outer ($r=r_H$) event horizon, corresponding to the roots of $\Delta=0$.

Taking into account the fact that the background solution is stationary and axisymmetric, one can decompose the perturbations into modes of  the form
\begin{eqnarray}
\label{defofpsi}
\psi(r,\theta) \, e^{-i \omega t + i m \varphi},
\end{eqnarray}
 where   $m$ is an integer and  $\omega$ is the frequency which turns out to be complex because of the damping of the modes due to gravitational radiation; $\psi$ is a function (parametrized by $\omega$ and $m$)  of  the radial and angular coordinates. 
 
In the case of spherical symmetry,  the perturbation equations, after decomposition into spherical harmonics,  reduce to  ordinary differential equations along the radial direction. By contrast, in the Kerr case one ends up with partial differential equations that depend on both $r$ and $\theta$. 
Remarkably, as shown by  Teukoslky \cite{Teukolsky:1973ha} upon using the Newman-Penrose formalism \cite{Newman:1961qr}, the perturbation equations can be expressed in a separable form. 
Indeed,  they are of  the form
\begin{eqnarray}
\label{Tequation}
{\cal O}(\omega,m) \psi = 0, 
\end{eqnarray}
where the second order differential operator ${\cal O}(\omega,m)$  can  be written as
\begin{eqnarray}
{\cal O}(\omega,m)  = {\cal O}_r(\omega,m) + {\cal O}_\theta(\omega,m)  \, ,
\end{eqnarray}
where ${\cal O}_r$ and ${\cal O}_\theta$ are second order differential operators involving respectively the variables $r$ and $\theta$ only.
As a consequence  $\psi$ can be written as a (sum of) products
$\psi(r,\theta)=R(r) S(\theta)$ where $R$ and $S$ satisfy separately ordinary differential equations. 
The details and the explicit equations  can be found in the original Teukolsky paper \cite{Teukolsky:1973ha}. 

The separability property renders the calculation of quasi-normal modes much more tractable. In this paper, we show that the equations for the tensor perturbations in DHOST theories that are known to admit a hairy Kerr solution remain separable. More precisely, these equations are exactly given by the classical equation  \eqref{Tequation} 
where now an effective ``source'' term appears and depends on the scalar perturbation, thus on the hair of the black hole solution. Remarkably, we find a parametrization of the modified scalar-tensor perturbations which allow us to compute the source term. 

\section{DHOST theories and stealth Kerr solution}

Let us now consider DHOST theories, which represent the most general family of covariant scalar-tensor theories propagating a single scalar degree of freedom.
We restrict our discussion to a subclass of DHOST theories which are shift and reflection symmetric $(\phi \to \phi + c , \, \phi \to - \phi)$ and whose tensor perturbations propagate at the speed of light \cite{Ezquiaga:2017ekz,Creminelli:2017sry,Baker:2017hug, Sakstein:2017xjx}. Their Lagrangian can be written in the form 
\begin{eqnarray}
{\cal L}  = K(X) + G(X)R + A_3(X)  L_3 + A_4(X)  L_4 +A_5(X)  L_5 \,, \nonumber \\ \label{DHOSTLagrangian} 
\end{eqnarray}
where $G$, $K$ and $A_I$ are functions of $X \equiv \phi_\mu \phi^\mu$, and
the Lagrangians $L_I$ are defined by
\begin{eqnarray}
L_3 \equiv \phi^\mu \phi^\nu \phi_{\mu\nu} \Box \phi \,, \;\;
L_4 \equiv \phi^\mu \phi_{\mu\rho} \phi_\nu \phi^{\nu\rho} \,, \;\;
L_5 \equiv (\phi^\mu \phi^\nu \phi_{\mu\nu})^2 \,.  \nonumber
\end{eqnarray} 
In our simplified notation, upper or lower indices on $\phi$ correspond to (covariant) derivatives, e.g. $\phi^\mu=\nabla^\mu\phi$ and  $\phi_{\mu\nu}=\nabla_\mu \nabla_\nu \phi$.
Due to the degeneracy conditions, which guarantee the presence of a single scalar degree of freedom, $A_4$ and $A_5$ are not free but depend on $G$ and $A_3$ as follows 
\cite{Langlois:2015cwa}
\begin{eqnarray}
A_4&=&  - A_3 + \frac{1}{8 G}\left(48 G_{X}^2 + 8 A_3 G_{X} X - A_3^2 X^2\right) \,, \label{deg1} \\
A_5 &=& \frac{A_3}{2 G}\left(4 G_X + A_3 X\right)\,.  \label{deg2}
\end{eqnarray} 
Furthermore, without loss of generality, since for our purposes we can consider the theory in vacuum, we can set $G=1$ by means of an $X$-dependent  conformal transformation \cite{Crisostomi:2016czh, Achour:2016rkg}.

In this subclass of theories, it is possible to construct stealth rotating black hole solutions, where the geometry is exactly Kerr (or Kerr-de Sitter if one adds an effective cosmological constant) and the 
scalar field $\phi$ is non-trivial \cite{Charmousis:2019vnf}. These solutions are characterised by a constant value of the kinetic term\footnote{Here we take $\phi^\mu$ to be timelike, should a spacelike $\phi^\mu$ be required, substitute $\mu^2 \to -\mu^2$ in what follows.} $X=X_0=-\mu^2$, and require that the following conditions are satisfied for the two independent functions $A_3$ and $K$  in the Lagrangian  \eqref{DHOSTLagrangian} (while $G=1$):
\begin{eqnarray}
A_3(X_0)=0 \,, \qquad
K(X_0)=0 \,, \qquad
K_X(X_0) = 0 \, .
\label{A3back}
\end{eqnarray}
Note that this implies, according to the degeneracy conditions (\ref{deg1}-\ref{deg2}), that $A_4(X_0)=0$ and $A_5(X_0)=0$ as well.

The scalar field $\phi$ can be obtained by exploiting the analogy with families of Kerr geodesics \cite{Carter:1968rr} and reads 
\begin{eqnarray}
\label{phisol}
\phi(t,r) = - \mu \, t + \varepsilon \mu \int \frac{\sqrt{R(r)}}{\Delta(r)} dr \, ,
\end{eqnarray}
where  
\begin{eqnarray}
R(r) \equiv 2M  r(r^2+a^2) \,,
\end{eqnarray}
and
$\varepsilon$ can take the values  $\pm 1$, corresponding to the two branches of the square root\footnote{Notice that, in absence of an effective cosmological constant (i.e. Kerr and not Kerr-dS metric), it is impossible to realise the merging of branches that characterises the solutions in \cite{Charmousis:2019vnf} and that provides a finite scalar field at both the event and cosmological horizons. We leave the study of perturbations around such solutions for future work.} in (\ref{phisol}).

\section{Equations for the perturbations}

We now  expand  the equations of motion for the scalar field and the metric to first order in perturbations, using 
\begin{eqnarray}
\label{perturbations}
g_{\mu\nu}= \overline{g}_{\mu\nu} + \delta g_{\mu\nu} , \quad
\phi = \overline{\phi} +  \, \delta \phi ,
\end{eqnarray} 
where barred quantities refer to the background whereas $\delta g_{\mu\nu}$ and $\delta \phi$ are perturbations. 
Interestingly,
 the leading order terms in the expansion of the equations
of motion simplify drastically, and one obtains, after  straightforward calculations, the equations
\begin{eqnarray}
&& \overline{\nabla}_\mu \left( \Xi \, \overline{\phi}^\mu \, \delta X \right)= 0 \,,  \label{eomX}\\
&&\delta G_{\mu\nu} = \delta T_{\mu\nu} \equiv \frac{1}{2} \Xi \, \overline{\phi}_\mu \overline{\phi}_\nu \delta X , \label{eomG}
\end{eqnarray}
where $\delta G_{\mu\nu}$ is the linearised Einstein tensor, $\delta T_{\mu\nu}$ is the effective source term associated with the scalar field and  $\delta X$ is the first order perturbation of the kinetic term.
We have also  introduced the notation 
\begin{eqnarray}
\Xi  \equiv  A_{3X}(X_0) \, {\cal E}_3 - 2 K_{XX}(X_0) , \label{Div=0} 
\end{eqnarray}
with 
\begin{eqnarray}
{\cal E}_3 &\equiv& ( \overline{\Box} \, \overline{\phi} )^2 - ( \overline{\phi}_{\mu\nu} )^2 \,.
\end{eqnarray}

When one replaces the background metric by  the Kerr solution (\ref{Kerr}) and the background scalar field by the expression \eqref{phisol},  the function $\Xi$  becomes 
\begin{eqnarray}
\label{expressionofA}
\Xi \; = \; 2 M \mu^2 a^2 \frac{3 \cos^2\theta + 1}{\rho^6} A_{3X}(X_0) - 2 K_{XX}(X_0) \, .
\end{eqnarray}

At this stage, let us make two comments. First, the equations for the perturbations do not involve the Lagrangians $L_4$ and $L_5$ in \eqref{DHOSTLagrangian}. Indeed, since the functions $A_3, A_4$ and $A_5$ vanish on the background, the quadratic expansion of the last three terms in the Lagrangian \eqref{DHOSTLagrangian} is given by
\begin{eqnarray}
\label{HOquadLag}
\delta X \left[ A_{3X}(X_0)  \delta L_3 + A_{4X}(X_0)  \delta L_4 + A_{5X}(X_0)  \delta L_5 \right] \, , 
\end{eqnarray}
where $\delta L_{I}$ is the first order perturbation of Lagrangian $L_I$. 
The latter can be easily rewritten as
\begin{eqnarray*}
L_3 = \frac{1}{2} \Box \phi \phi^\mu \partial_\mu X \,, \quad
L_4=\frac{1}{4} \partial_\mu X \partial^\mu X \,, \quad
L_5=\frac{1}{4} (\phi \phi^\mu \partial_\mu X )^2 \,,
\end{eqnarray*}
hence, we see immediately that  $\delta L_4 = \delta L_5 =0$ at  linear order. As a consequence, the quadratic Lagrangian \eqref{HOquadLag} reduces to
\begin{eqnarray}
\frac{1}{4} A_{3X}(X_0) \overline{\Box} \, \overline{\phi} \,  \overline{\phi}^\mu \partial_\mu (\delta X^2) .
\end{eqnarray}

Second, we see that using the variable $\delta X$, which is related to the original perturbations  \eqref{perturbations} 
through the equation
\begin{eqnarray}
\label{deltaXdef}
\delta X = 2 \overline{\phi}^\mu \partial_\mu (\delta \phi) -  \overline{\phi}^\mu \overline{\phi}^\nu \delta g_{\mu\nu}\,,
\end{eqnarray}
considerably simplifies the dynamics. 
Indeed, $\delta X$ is totally decoupled
from the Einstein tensor perturbations and, as we are going to see in the next section, its equation can be solved explicitly. As a consequence, the
equations for the tensor perturbations reduce to the linearized  Einstein equations supplemented with a source term which depends on $\delta X$. Therefore, the Teukolsky equations \eqref{Tequation} for the Newman-Penrose coefficients $\psi$ is exactly the same as in general relativity with a source term that can be explicitly computed.

 \section{Solution for the scalar kinetic density perturbation $\delta X$}

We now solve equation \eqref{eomX} for $\delta X$.
This equation is first order in $\delta X$ and therefore can be easily integrated. 

First, we write it as follows
\begin{eqnarray}
\label{Div=0}
\partial_\mu \left( \sqrt{-g}\, \Xi \, \phi^\mu \delta X\right) = 0 \, ,
\end{eqnarray}
where $\Xi$ is given by \eqref{expressionofA},  and the determinant of the metric $g$ and the non-vanishing components of $\phi^\mu$ are given by
\begin{eqnarray}
\sqrt{-g} & = & \rho^2 \sin\theta, \\
\phi^t & = & \frac{\mu}{\Delta \, \rho^2} \left[ (r^2+a^2)^2 - a^2 \Delta \sin^2\theta \right] ,\\
\phi^r & = & \varepsilon \mu \frac{\sqrt{R}}{\rho^2} , \quad \phi^\varphi = 2M a \mu \frac{r}{\Delta \, \rho^2} \,. 
\end{eqnarray}
As $\phi$ does not depend on $\theta$ (in the Kerr geometry), the component
$\phi^\theta = g^{\theta \theta} \phi_\theta $ vanishes. Substituting these expressions into \eqref{Div=0} leads to the very
simple form
\begin{eqnarray}
 \varepsilon {\Delta}  {\partial}_r(\sqrt{R} \chi) 
+  \left[ {(r^2+a^2)^2}- a^2 {\Delta} \sin^2 \theta \right] \partial_t \chi + 2Ma r \partial_\varphi \chi = 0 \,, \nonumber 
\end{eqnarray}
where we have introduced the variable $\chi \equiv \Xi \delta X$ for simplicity. We decompose the solution into modes 
\begin{eqnarray}
\label{funcchi0}
\chi = \sum_{m} \int d \omega \, \chi_{m,\omega}(r,\theta) e^{-i\omega t + i m \varphi} \, ,
\end{eqnarray}
and we easily obtain the general solution for each  mode
\begin{eqnarray}
\label{funcchi}
&&\chi_{m,\omega}(r,\theta) = \\
&&\qquad \frac{C_{m,\omega}(\theta)}{\sqrt{R(r)}} \exp\left[ i \varepsilon  \left(-\omega I(r) - \omega \sin^2\theta J(r) + m K(r)\right) \right] \,, \nonumber
\end{eqnarray}
where $C_{m,\omega}$ is, at this stage, an arbitrary function of $\theta$ (parametrized by $\omega$ and $m$) and
\begin{eqnarray}
I(r) & \equiv & - \int dr \, \frac{(r^2+a^2)^2}{\Delta(r) \sqrt{R(r)}} \,,\label{IntI}\\ 
J(r) & \equiv & \int  dr \, \frac{a^2}{\sqrt{R(r)}} \,, \label{IntJ}  \\ 
K(r) & \equiv & - \int dr \, \frac{2Ma r}{\Delta(r) \sqrt{R(r)}} \,. \label{IntK} 
\end{eqnarray}
Thus, the components $\delta X_{m,\omega}$ of the perturbation $\delta X$ are immediately given by  $\delta X_{m,\omega}=\chi_{m,\omega}/\Xi$.

\section{Behaviour of the solution at the boundaries}

Now, let us study the regularity of this solution. We start analyzing the behavior of the modes \eqref{funcchi} when $r$ approches 
the horizon $r_H$, where $\Delta$ vanishes. 
In this limit, one finds a divergence in the integrals $I$ and $K$. However, the  Boyer-Lindquist coordinates also become singular at the horizon and one must use, instead, well-behaved coordinates such as  the ingoing Eddington-Finkelstein-like coordinates  for Kerr, which we denote $\{v, r,\theta,\tilde\varphi \}$. The coordinate transformation is given by 
\begin{eqnarray}
v \equiv t + \int dr \, \frac{r^2+a^2}{\Delta(r)} \, , \quad
\tilde\varphi \equiv \varphi + a \int \frac{dr}{\Delta(r)} \, .
\end{eqnarray}
As a consequence, the combination $(t+\varepsilon I)$ which appears in \eqref{funcchi0} can be rewritten as 
\begin{eqnarray}
t+\varepsilon I=v-\int dr \frac{\rho_0^{3}\left(\varepsilon\rho_0+\sqrt{2Mr}\right)}{\sqrt{2Mr}(\rho_0-\sqrt{2Mr})(\rho_0+ \sqrt{2Mr})}
\end{eqnarray}
with $\rho_0(r)\equiv (r^2+a^2)^{1/2}$, which shows that the singularity disappears for the branch $\varepsilon=-1$. Similarly, we have 
\begin{eqnarray}
\varphi+\varepsilon K=\tilde\varphi -a\int dr  \frac{\sqrt{2Mr}\left(\varepsilon \sqrt{2Mr}+\rho_0\right)}{(\rho_0-\sqrt{2Mr})(\rho_0+ \sqrt{2Mr})}
\end{eqnarray}
and the singularity also disappears when $\varepsilon=-1$. This is in complete accord with the background solution as $\varepsilon=-1$ corresponds to the branch where the scalar field is regular at the event horizon.

One can also examine  the behavior of the modes \eqref{funcchi} at large distances. 
Clearly, the functions $J$ and $K$ converge when $r$ tends to infinity, whereas $I$ diverges 
according to $I(r)\sim - r^{3/2}$. Taking into account that the  frequency $\omega$ contains an imaginary part, which depends on the typical damping time,  Im$(\omega) \equiv - 1/\tau <0$,
one finds that the mode is suppressed at large distances in the branch 
$\varepsilon=-1$ (whereas it diverges for $\varepsilon=1$). 
 
In summary, the explicit solution for $\delta X$ that we have obtained appear well-behaved both at the horizon and at spatial infinity in the branch $\varepsilon=-1$.
Note that the solution for each mode depends on an arbitrary  function $C_{m,\omega}$ of the angular variable $\theta$, which is in principle determined by the initial conditions.

\section{Discussion}
In this work, we have studied the perturbations of a stealth rotating black hole solution in a subclass of DHOST theories.  The background solution is characterized by a constant value of the scalar field  kinetic term $X$ and its perturbation $\delta X$ fully describes how the tensor modes are modified in this model. Indeed we have found that the equations of motion for the perturbations of the geometry can be reformulated as  linearized Einstein's equation with  a source term proportional to $\delta X \bar\phi_\mu \bar\phi_\nu$, in contrast with the GR result. 
In parallel, the quantity $\delta X$ obeys an equation that does not involve the tensor modes and can thus be solved independently.  We have written the general solution of this equation as a superposition of  modes and studied their behaviour at the black hole horizon and at spatial infinity, finding that they are well-behaved only for one branch of the background solutions.

Given a solution for $\delta X$, one can then compute the source term in the Teukolsky equation for the Newman-Penrose
 variables $\psi$.
Contrary to general relativity, the equation for $\psi$ is no longer homogeneous.
The situation  thus appears  analogous to the simpler case of a vibrating string obeying a dissipative wave equation with a source term, i.e. of the form
\begin{eqnarray}
\frac{\partial^2 \psi }{\partial x^2} - \frac{\partial^2 \psi }{\partial t^2} - \frac{2}{\tau} \frac{\partial \psi}{\partial t}= S(x,t).
\end{eqnarray}
Assuming the boundary conditions $\psi(0,t)=\psi(L,t)=0$ (and for the source as well), $\psi$ can be decomposed as
\begin{eqnarray}
\label{stringQNM}
\psi(x,t) = \sum_{n} \; A_n \; \sin(n\pi x/L) \; e^{-i \omega_n t},
\end{eqnarray}
where the complex frequencies $\omega_n$, are given by
\begin{eqnarray}
\label{stringmodes}
\omega_n \tau \equiv - i \pm \sqrt{n^2\frac{\pi^2\tau^2}{L^2}-1} \,.
\end{eqnarray}
These discrete  $\omega_n$'s are analogous to the black hole QNM. In the absence of source, the full solution is given by (\ref{stringQNM}) where the coefficients $A_n$ are fixed by the initial conditions at $t=0$.

When a source term is present, it is convenient to   rewrite it in the form 
\begin{eqnarray}
S(x,t)=\sum_n \int d\omega \, \sin (n\pi x/L) \, e^{-i\omega t} \, \hat{S}(n,\omega),
\end{eqnarray}
then, the solution of the wave equation is formally
given by
\begin{eqnarray}
\psi(x,t) = - \sum_n \int d\omega  \, \frac{\hat{S}(n,\omega)  \sin (n\pi x/L) \, e^{-i\omega t} }{n^2\pi^2/ L^2 - \omega^2 - 2i\omega/\tau}\,.
\end{eqnarray}
Such an analysis   for the QNMs of  a GR Kerr black hole in presence of a source term due to some matter around the black hole,    can be found in   \cite{Detweiler:1977gy}. 

In the specific case of the stealth black hole, one could proceed along the same lines. This would however require a more precise description of the source term, i.e. identify which of the solutions for $\delta X$ are physically relevant and how they can be generated in some physical process, analysis which goes beyond the scope of the present work.

 To conclude, our analysis is a first direct attempt to tackle perturbations of non trivial hairy rotating black holes.  It is still not clear whether this model can be seen as a viable alternative to GR or needs to be further restricted. 
There could  also be some issues concerning the  validity of hairy stealth solutions  from an effective field theory point of view (see e.g. discussions in \cite{Babichev:2018uiw} and very recently  \cite{deRham:2019gha}). These aspects should be explored further in the future (see for example \cite{Endlich:2010hf}).

\section*{Acknowledgements}
We thank Stas Babak, Eugeny Babichev, Enrico Barausse, Brando Bellazzini, Pedro Ferreira, Shinji Mukohyama, Francesco Nitti, George Pappas and Nick Stergioulas for useful discussions.
CC thanks  the Laboratory of Astronomy of AUTh in Thessaloniki for hospitality.
CC and KN acknowledge support from the CNRS project 80PRIME.
MC is supported by the Labex P2IO.


\bibliography{QNM}
\bibliographystyle{Biblio}

\end{document}